\documentclass[12pt]{article}
\usepackage{arxiv}

\usepackage[utf8]{inputenc} %
\usepackage[T1]{fontenc}    %
\usepackage{booktabs}       %
\usepackage{amsfonts}       %
\usepackage{nicefrac}       %
\usepackage{microtype}      %

\usepackage{url}
\usepackage{xcolor}         %
\usepackage[inline]{enumitem} %

\usepackage[numbers,sort&compress]{natbib}
\usepackage{doi}
\usepackage{algorithm,algpseudocode}
\usepackage{amsmath}
\usepackage{amssymb}
\usepackage{mathtools}
\usepackage{amsthm}
\usepackage{multirow, array, threeparttable}
\usepackage[font=normalsize,skip=4pt]{caption}
\usepackage{subcaption}

\usepackage{graphicx}   %
\hypersetup{
    colorlinks,
    linkcolor={blue!60!black},
    citecolor={red!60!black},
    urlcolor={blue!80!black}
}
\usepackage[capitalize,nameinlink]{cleveref}

\makeatletter
\renewcommand{\paragraph}{%
  \@startsection{paragraph}{4}{\z@}%
  {0.25\baselineskip}  %
  {-0.25em}            %
  {\normalfont\normalsize\bfseries}%
}
\makeatother

\usepackage{enumitem}

\algnewcommand\algorithmicinput{\textbf{Input:}}
\algnewcommand\algorithmicoutput{\textbf{Output:}}
\algnewcommand\algorithmicnote{\textbf{Note:}}
\algnewcommand\Input{\item[\algorithmicinput]}%
\algnewcommand\Output{\item[\algorithmicoutput]}%
\algnewcommand\Note{\item[\algorithmicnote]}%

\newcommand{\diff}{\,\mathrm{d}} %

\title{Achieving high-performance polarization-independent nonreciprocal thermal radiation with pattern-free heterostructures}

\author{
{\hspace{1mm}Bach Do$^\star$}\\
	University of Houston\\
	\texttt{bdo3@uh.edu} \\
        \And
{\hspace{1mm}Bardia Nabavi$^\star$}\\
	University of Houston\\
	\texttt{bnabavi@uh.edu} \\
        \And
{\hspace{1mm}Sina Jafari Ghalekohneh}\\
	University of Houston\\
	\texttt{sinajafari@uh.edu} \\
        \And
{\hspace{1mm}Taiwo Adebiyi} \\
	University of Houston\\
	\texttt{taadebiyi2@uh.edu} \\
        \And
{\hspace{1mm}Bo Zhao$^*$} \\
	University of Houston\\
	\texttt{bzhao8@uh.edu} \\
        \And
 {\hspace{1mm}Ruda Zhang$^*$} \\
	University of Houston\\
	\texttt{rudaz@uh.edu} \\
}

\date{}

\renewcommand{\headeright}{ }
\renewcommand{\shorttitle}{Pattern-free Polarization-independent Nonreciprocity}

\begin{document}
\maketitle

\begin{abstract}
Many advanced energy harvesting technologies rely on advanced control of thermal emission. Recently, it has been shown that the emissivity and absorptivity of thermal emitters can be controlled independently in nonreciprocal emitters. While significant progress has been made in engineering these nonreciprocal thermal emitters, realizing a highly efficient, pattern-free emitter capable of supporting dual-polarization nonreciprocal emission remains a challenging task.
Existing solutions are largely based on metamaterials and exhibit polarization-dependent behavior.
This work proposes pattern-free multilayer heterostructures combining magneto-optical and magnetic Weyl semimetal materials and systematically evaluates their nonreciprocal emission performance for $p$- and $s$-polarized waves. %
The findings show that omnidirectional polarization-independent nonreciprocity can be achieved utilizing multilayer structures with different magnetization directions that do not follow simple vector summation.
To further enhance the performance, Pareto optimization is employed to tune the key design parameters, enabling the maximization of nonreciprocal thermal emission in a given wavelength range.
This approach offers a versatile strategy for designing high-performance thermal emitters tailored for multi-objective optical functionalities.
\end{abstract}


\def\thefootnote{$\star$}\footnotetext{Equal contribution.}\def\thefootnote{\arabic{footnote}}
\def\thefootnote{$*$}\footnotetext{Corresponding authors.}\def\thefootnote{\arabic{footnote}}

\section{Introduction} 
Thermal emitters are essential components in a wide range of radiative energy-harvesting technologies including solar cells, thermophotovoltaics, and solar thermophotovoltaics, where precise control of thermal radiation plays a critical role in maximizing energy conversion efficiency \cite{Greffet2002, FanS2017, Baranov2019,jafari2022nonreciprocal}. For reciprocal emitters, thermal emission obeys Kirchhoff’s law of thermal radiation, which enforces equality between directional spectral emissivity and absorptivity \cite{ planck1914theory,kirchhoff1978verhaltnis}. This inherent reciprocity, however,  imposes fundamental limits on the performance of thermal emitters, making energy losses unavoidable \cite{zhang2022nonreciprocal,YangS2024}.  

To overcome these limitations, nonreciprocal thermal emitter technologies that violate Kirchhoff’s law \cite{snyder1998thermodynamic,zhu2014near,zhao2019near,zhang2022nonreciprocal,shayegan2023direct} have emerged as a promising pathway  \cite{jafari2022nonreciprocal,ghalekohneh2025perfectheatrectificationcirculation}. By breaking reciprocity, one can independently control the directional spectral emissivity and absorptivity, thus minimizing radiative losses while maximizing useful emission. Such emitters improve emission in targeted channels, offering significant potential for advanced energy conversion and radiative heat flow control \cite{jafari2022nonreciprocal, ghalekohneh2025perfectheatrectificationcirculation}. 

Nonreciprocity in thermal systems can be realized through nonlinearity \cite{khandekar2015radiative}, time-variant modulation \cite{ghanekar2022violation}, or magnetic effects \cite{shayegan2023direct}. Among these, magneto-optical (MO) materials, such as doped semiconductors \cite{zvezdin1997modern,shayegan2022nonreciprocal}, which require external magnetic fields, and magnetic Weyl semimetals, which inherently break time-reversal symmetry \cite{kotov2018giant,ZhaoB2020}, have emerged as a particular route due to their unique characteristics. In MO materials, nonreciprocity originates from the magneto-optical Kerr effect (MOKE), which renders the permittivity tensor asymmetric ($\boldsymbol{\varepsilon} \neq \boldsymbol{\varepsilon}^\intercal$) under an external magnetic field \cite{ qiu2000surface, madelung2004semiconductors}. This asymmetry introduces polarization-dependent phase and amplitude changes upon reflection, leading to nonreciprocal thermal emission \cite{zhang2022nonreciprocal}. In comparison, the momentum separation between Weyl nodes in magnetic Weyl semimetals emulates the effect of an external magnetic field, enabling intrinsic nonreciprocity without applied magnetic bias \cite{armitage2018weyl, ZhaoB2020, guo2023light}.

\begin{figure*}[t]
	\centering
    \includegraphics[scale=0.68]{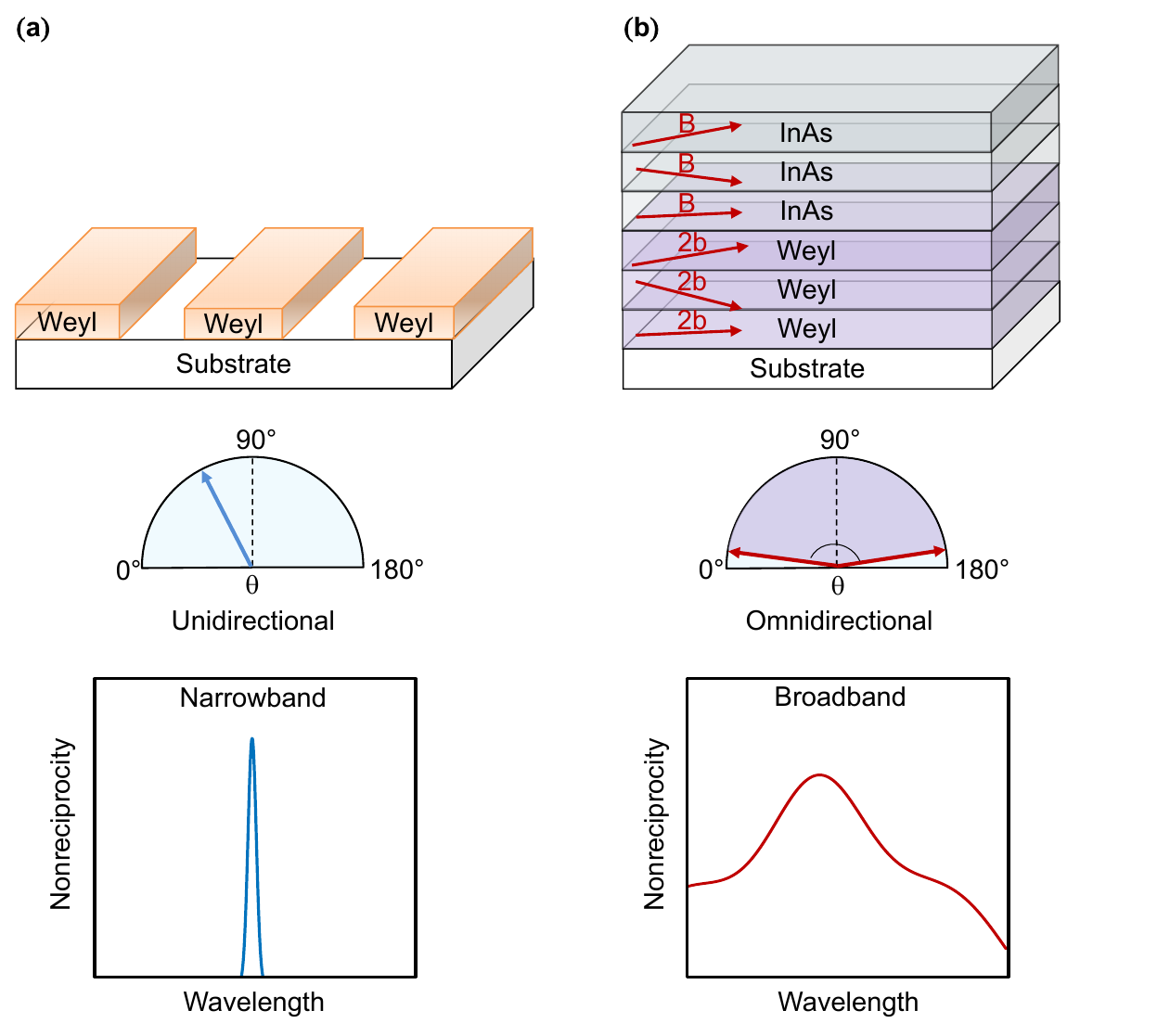}
	\caption{Design of dual-polarized nonreciprocal thermal emitters. (a) Patterned structures enables narrowband directional nonreciprocity by triggering bound state in continuum (BIC) modes \cite{wu2025mid,fang2024dual, xia2022circular}. (b) Our proposed design realizes dual-polarized broadband omnidirectional nonreciprocal emission by optimizing the 
magnetization configuration in a multilayer heterostructure composed of magneto-optical materials and Weyl semimetals, without requiring intricate patterns or nanostructures. $\theta$ denotes polar angle and nonreciprocity is illustrated as contrast between emissivity and absorptivity in comparison to wavelength.
}
	\label{fig:FOM}
\end{figure*}

Due to these appealing optical properties, MO materials and magnetic Weyl semimetals, have been widely studied for nonreciprocal thermal emitters \cite{YangS2024}. In these studies, however, most achieved nonreciprocal effects are limited to  $p$-polarized waves (magnetic field perpendicular to the plane of incidence), whereas nonreciprocity is generally absent for $s$-polarized waves (electric field perpendicular to the plane of incidence)\cite{fang2023dual}. Since thermal radiation is inherently incoherent and unpolarized, realizing dual-polarization nonreciprocity is crucial.

 A few recent studies have demonstrated dual-polarization nonreciprocity using complex structures (\textbf{Figure}~\ref{fig:FOM}). For example,  Xia \textit{et al.}~\cite{xia2022circular} demonstrated both experimentally and theoretically that a sub-wavelength dielectric Mie resonator on top of an MO material can produce anomalous $s$-polarized nonreciprocity through circular displacement currents associated with magnetic dipole and quadrupole modes in the transverse and longitudinal configurations. Fang \textit{et al.}~\cite{fang2024dual} employed patterned silicon nanopore arrays to nearly fully violate Kirchhoff’s law under both polarizations by applying a $4$ T magnetic field at a fixed polar angle, enabled by the coexistence of guided-mode and cavity-mode resonances.
More recently, Wu \textit{et al.}~\cite{wu2025mid} designed an emitter with a single layer of Weyl semimetal, achieving dual-polarized nonreciprocal emission through azimuthal angle tuning in a narrowband regime. Although significant progress have been made, reported designs remain limited to narrow bandwidths, unidirectional operation, or reliance on complex nanofabrication techniques such as photonic crystals and subwavelength patterning (\textbf{Figure}~\ref{fig:FOM}(a)). Achieving broadband, omnidirectional, and dual-polarized nonreciprocity in a fabrication-friendly platform remains an open challenge. 

In this work, we propose a texture-free multilayer heterostructure composed of indium arsenide (InAs) and magnetic Weyl semimetal to realize high-performance, polarization-independent nonreciprocal thermal radiation (\textbf{Figure}~\ref{fig:FOM}(b)). Our design leverages the interplay between gradient doped structures, and MOKE in stacked layers to break reciprocity omnidirectionally for both $p$- and $s$-polarizations without requiring surface patterning. The main challenge lies in optimizing multiple coupled parameters simultaneously, including magnetization strength and orientation, carrier concentration, and geometric layer properties. To address this, we employ a Pareto optimization framework \cite{Deb2002,ngatchou2005pareto} to identify configurations that maximize dual-polarization nonreciprocity in the infrared regime. Our approach provides a fabrication-friendly alternative for achieving dual-polarization nonreciprocal thermal emission, eliminating the need for intricate nanostructuring. Our study indicates substantial potential for designing nonreciprocal thermal emitters that enhance nonreciprocal contrast, particularly for $s$-polarized and dual-polarized waves. Distinct optimized designs can be achieved for both $s$- and $p$-polarized emitters, reflecting their different underlying physical behaviors. Furthermore, the dominance of $p$-polarized nonreciprocity in the unpolarized case reveals a new pathway for defining and engineering dual-polarization nonreciprocity. These insights open new avenues for future research on high-performance nonreciprocal thermal emitters.

\section{Absorptivity and emissivity of nonreciprocal emitters with polarization conversion}
\label{Sec2}
\begin{figure*}[t]
	\centering
	\includegraphics[scale=0.6]{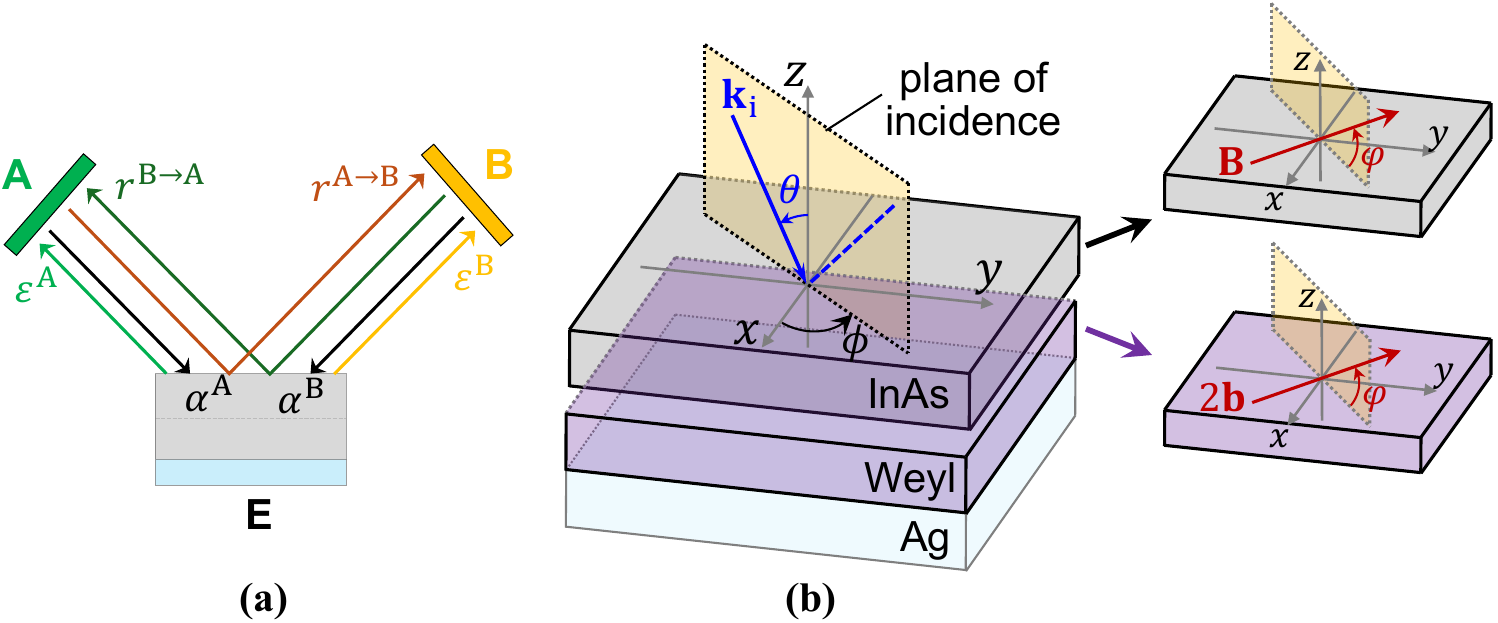}
	\caption{
	(a) Schematic of radiative heat exchange between two blackbodies (A and B) and an emitter (E) under thermal equilibrium. (b) A pattern-free heterostructure consisting of InAs layers on top of Weyl semimetal layers on an Ag substrate.  The incident wavevector $\mathbf{k}_{\mathrm{i}}$ is characterized by an azimuthal 
angle $\phi$ and a polar angle $\theta$. The azimuthal angle $\phi$ is the angle between 
$\mathbf{k}_{\mathrm{i}}$
and the $x$-axis and varies within $[-180^{\circ}, 180^{\circ}]$. The polar 
angle $\theta$ is defined as the angle between $\mathbf{k}_{\mathrm{i}}$ and the 
$z$-axis, ranging from $[0^{\circ}, 90^{\circ}]$. 
The angle $\varphi$ specifies the orientation of the external magnetic field 
$\mathbf{B}$ (or the momentum-separation vector $2\mathbf{b}$ in a Weyl semimetal) 
relative to the plane of incidence, and takes values within 
$[-180^{\circ}, 180^{\circ}]$. Counterclockwise direction corresponds to 
positive values.}
	\label{fig:EmittersStruct}
\end{figure*}

Consider a system in thermal equilibrium consisting of an opaque thermal emitter E exchanging energy with two blackbodies, A and B, as depicted in \textbf{Figure}~\ref{fig:EmittersStruct}(a) \cite{nabavi2025high}.  Level of nonreciprcity $\eta_{j}$ is represented as the contrast between the spectral directional emissivity and absorptivity of the emitter \cite{zhu2014near}:
\begin{equation}
\eta_j (\theta,\phi,\lambda) = \varepsilon_j^{\mathrm{A}}(\theta,\phi,\lambda)
   - \alpha_j^{\mathrm{A}}(\theta,\phi,\lambda) \label{eq:1}
\end{equation}
where $\varepsilon_j^{\mathrm{A}} $ and $\alpha_j^{\mathrm{A}}$ are spectral directional emissivity and absorptivity, respectively, for $j$-polarized light ($j \in \{s, p\}$). Here, $\theta$ is the polar angle, $\phi$ is the azimuthal angle, and $\lambda$ is the wavelength. To understand general nonreciprocal thermal emitters with polarization conversion, we discuss the energy balance in these systems. The emission from blackbody A is either absorbed by the emitter E or reflected toward blackbody B. Since $s$- and $p$-polarized waves behave identically in a blackbody and share the emitted energy equally, the outgoing energy can be written separately for each polarization:

\begin{equation}
\alpha_j^{\mathrm{A}}  (\theta,\phi,\lambda) + r_j^{\mathrm{A} \rightarrow \mathrm{B}} (\theta,\phi,\lambda) = 1, \label{eq:2}
\end{equation}
where $r_{j}^{\mathrm{A}\rightarrow \mathrm{B} }$ is spectral directional reflectivity, for $j$-polarized light from A to B. Meanwhile, blackbody A absorbs all energy directed to it ($\alpha=1$) and the amount is the same as it emits as required by thermodynamics. The energy absorbed is from the emitter $\varepsilon_j^{\mathrm{A}} $ or reflected from the emitter $r_{j}^{\mathrm{B}\rightarrow \mathrm{A} }$. Due to the MOKE, the polarization of the incident wave changes upon interacting with the emitter. An incident $p$-polarized wave emitted from B is partially converted into an $s$-polarized component, while an $s$-polarized wave undergoes a corresponding conversion into a $p$-polarized component. As a result, when considering the absorbed energy by A, the $s$-polarized channel must be included when evaluating the $p$-polarized response, and the $p$-polarized channel must likewise be included when evaluating the $s$-polarized response. Consequently, the incoming $p$- and $s$-polarized waves cannot be treated independently in energy balance analysis and must be considered simultaneously
\begin{equation}
\begin{aligned}
&\varepsilon_p^{\mathrm{A}}(\theta,\phi,\lambda) 
+ \varepsilon_s^{\mathrm{A}}(\theta,\phi,\lambda)\\
&+|R_{ss}^{\mathrm{B} \rightarrow \mathrm{A}} (\theta,-\phi,\lambda)|^2+|R_{sp}^{\mathrm{B} \rightarrow \mathrm{A}} (\theta,-\phi,\lambda)|^2+|R_{pp}^{\mathrm{B} \rightarrow \mathrm{A}} (\theta,-\phi,\lambda)|^2 
+|R_{ps}^{\mathrm{B} \rightarrow \mathrm{A}} (\theta,-\phi,\lambda)|^2 = 2.
\label{eq:3}
\end{aligned}
\end{equation}
where $r_{j}^{\mathrm{A} \rightarrow \mathrm{B}}(\theta,\phi,\lambda)=|R_{jj}^{\mathrm{A} \rightarrow \mathrm{B}}(\theta,\phi,\lambda)|^2+|R_{ji}^{\mathrm{A} \rightarrow \mathrm{B}}(\theta,\phi,\lambda)|^2$ . $R_{ji}$ and $R_{jj}$ are cross- and co-polarized reflection coefficients ($j,i \in \{s, p\}$).  
As observed above, the $p$- and $s$-polarized waves are coupled, and therefore the $j$-polarized nonreciprocal contrasts defined in \Cref{eq:1} cannot be evaluated independently in contrast to earlier studies in which polarization coupling is absent \cite{zhu2014near}. To address this complexity, we introduce an additional law that enables the separated study of each polarization channel. We implicate adjoint Kirchhoff's law which shows that an adjoint system can be designed in which its absorptivity is equal to emissivity of the original system \cite{GuoC2022}
\begin{equation}
\varepsilon_j^{\mathrm{A}}(\theta,\phi,\lambda)=\alpha_j^{\mathrm{B} } (\theta,-\phi, \,\lambda),\label{eq:4} 
\end{equation}
and contrast can be reformulated by comparing the $j$-polarized absorptivity by plugging $\alpha_j^{\mathrm{B} } (\theta,-\phi,\lambda) $ in \Cref{eq:4} into \Cref{eq:1} yields
\begin{equation}
\begin{split}
\eta_j(\theta,\phi,\lambda)=\alpha_j^{\mathrm{B} }(\theta,-\phi,\lambda) -\alpha_j^{\mathrm{A} }(\theta,\phi,\lambda)  
=r_j^{\mathrm{B} \rightarrow\mathrm{A} }(\theta,-\phi,\lambda) -r_j^{\mathrm{A} \rightarrow\mathrm{B}}(\theta,\phi,\lambda),   
\label{eq:5}
\end{split}
\end{equation}
which greatly simplifies the calculation.

In this study, we employ doped InAs \cite{DongJ2021, liu2023broadband} and magnetic Weyl semimetal \cite{pajovic2020intrinsic, picardi2024nonreciprocity}, to achieve nonreciprocity in planar, pattern-free structures (\Cref{fig:EmittersStruct}(b)). We first study single layer doped InAs (or magnetic Weyl semimetal) on a reflective silver (Ag) substrate with arbitrary magnetization directions to study the effect of thin layer magnetization. Subsequently, the emitter is optimized to multilayer structures  to simultaneously maximize dual-polarized nonreciprocal contrasts.

Consider a configuration in which the InAs sample is oriented such that the plane of incidence is along $xz$ plane (i.e., $\phi = 0^\circ$) and applied magnetic field $\mathbf{B}$ is along the $y$-direction. Therefore, the permittivity tensor of InAs can be expressed as \cite{zhu2014near}
\begin{equation}
\boldsymbol{\varepsilon} =
\begin{bmatrix}
\varepsilon_{xx} & 0 & \varepsilon_{xz} \\
0 & \varepsilon_{yy} & 0 \\
\varepsilon_{zx} & 0 & \varepsilon_{zz}
\end{bmatrix}. \label{eq:6}
\end{equation}
The permittivity components follow the Drude, as
\begin{align}
\varepsilon_{xx} &= \varepsilon_{yy} = \varepsilon_\infty - \frac{\omega_p^2 (\omega + i \Gamma)}{\omega [(\omega + i \Gamma)^2 - \omega_c^2]}, \label{eq:7} \\
\varepsilon_{xz} &= - \varepsilon_{zx} = \frac{i \omega_p^2 \omega_c}{\omega [(\omega + i \Gamma)^2 - \omega_c^2]}, \label{eq:8} \\
\varepsilon_{zz} &= \varepsilon_\infty - \frac{\omega_p^2}{\omega (\omega + i \Gamma)},
\label{eq:9} 
\end{align}
where $\varepsilon_\infty$ is the high-frequency permittivity, $\omega$ is the angular frequency, $\Gamma$ is the damping rate, $\omega_p = \sqrt{n_e e^2 / (m^* \varepsilon_0)}$ is the plasma frequency, and $\omega_c = eB / m^*$ is the cyclotron frequency.
Here, $n_e$ is the carrier concentration, $e$ is the elementary charge, $m^{*}$ is the effective electron mass, $\varepsilon_{0}$ is the vacuum permittivity, and $B$ is the external magnetic field in the Drude model~\cite{XuJ2021}. The ratio $\kappa=\varepsilon_{xz} / \varepsilon_{xx}$ evaluates the degree of nonreciprocity \cite{shayegan2023direct}. This permittivity tensor is specifically valid when the external magnetic field is aligned along the $y$-direction. For cases with an arbitrary magnetization direction, the general permittivity tensor can be obtained by applying an appropriate coordinate transformation matrix \cite{yeh1990optical}. Building on this formulation, we developed a custom $4\times4$ scattering matrix framework to compute the nonreciprocal response of our multilayer structures under general magnetic field orientations, where all in-plane tensor components may contribute to the optical behavior.  Nonreciprocity is maximized in the epsilon-near-zero (ENZ) region, where $\kappa$ attains large values as $\varepsilon_{xx} \rightarrow 0$ while $\varepsilon_{xz} \neq 0$ \cite{Campione2015, Halterman2018, Kinsey2019, XuJ2021}. The ENZ region can be tuned through the doping level, allowing control of nonreciprocity across different wavelength ranges \cite{JafariGhalekohne2024}.

Magnetic Weyl semimetals inherently break time-reversal symmetry due to the anomalous Hall effect~\cite{tsurimaki2020large, okamura2020giant}. Their internal magnetization arises from Weyl node momentum separation $2\textbf{b}$~which mimics the applied field \cite{nandy2017chiral, armitage2018weyl}. If $2\textbf{b}$ is aligned along the $y$-direction, then the permittivity of the magnetic Weyl semimetal can also be described as in \Cref{eq:8}, where~\cite{chang2015chiral, ZhaoB2020}
\begin{equation}
\varepsilon_{xx} = \varepsilon_{yy} = \varepsilon_{zz} = \varepsilon_b + \frac{i\sigma}{\omega \varepsilon_0},
\label{eq:10}
\end{equation}
\begin{equation}
\varepsilon_{xz} = -\varepsilon_{zx} = i \frac{b e^2}{2\pi^2 \hbar \omega}.
\label{eq:11}
\end{equation}
Here, $\varepsilon_b$ is the background permittivity, and $\sigma$ is the bulk conductivity derived from the two-band model~\cite{chang2015chiral, ZhaoB2020}, as
\begin{equation}
    \sigma = \frac{\varepsilon_0 r_s g E_F}{6\hbar} \, \Omega \, 
G\left( \frac{E_F \Omega}{2} \right)  
+ i \frac{\varepsilon_0 r_s g E_F}{6\pi \hbar} 
\left\{ \frac{4}{\Omega} \left[ 1 + \frac{\pi^2}{3} \left( \frac{k_B T}{E_F} \right)^2 \right] 
+ 8 \Omega \int_0^{\xi_c} 
\frac{ G(E_F \xi) - G(E_F \Omega / 2) }{ \Omega^2 - 4\xi^2 } \, \xi \, \diff \xi \right\},
\label{eq:12}
\end{equation}
where $r_s = e^2 / \left( 4\pi \varepsilon_0 \hbar v_F \right)$ is the effective fine-structure constant, $g$ is the number of Weyl nodes, $E_F$ is the Fermi level, $\hbar$ is the reduced Planck constant, and $v_F$ is the Fermi velocity. $\Omega = \hbar \left( \omega + i \tau^{-1} \right)$ is the normalized complex frequency, and $\tau$ is the scattering time. $G(E) = n(-E) - n(E)$, where $n(E)$ is the Fermi--Dirac distribution. $k_B$ is the Boltzmann constant, and $\xi_c = E_C / E_F$ is the normalized cutoff energy, with $E_C$ being the cutoff energy beyond which the band dispersion is no longer linear. Similar to the carrier concentration $n_e$ in InAs, tuning the Fermi level $E_F$ allows control over the spectral response to shift the ENZ point~\cite{Halterman2018}.

\section{Nonreciprocity caused by magneto-optical Kerr effect}

Nonreciprocity in thin-film structures of magneto-optical materials arises from the MOKE, which can be classified into three categories depending on the magnetization direction \cite{qiu2000surface,madelung2004semiconductors}: Voigt or transverse (magnetic field along the \textit{y}-direction, $\varphi = 90^\circ$), longitudinal (magnetic field along the \textit{x}-direction, $\varphi = 0^\circ$), and polar MOKE configuration (magnetic field perpendicular to the film surface, along the \textit{z}-direction), where the plane of incidence lies in the \textit{xz} plane and $\varphi$ (see \Cref{fig:EmittersStruct}(b)) is the angle between the magnetization direction and the plane of incidence. For planar structures, nonreciprocal emission is typically observed only for $p$-polarized waves in the Voigt configuration, where the polarization of the incident and reflected waves remains unchanged \cite{shayegan2022nonreciprocal}. In this case, the electric field of the $p$-polarized wave lies in the plane of incidence (the $xz$ plane), while the magnetic field is perpendicular to this plane (along the $y$-direction). 
The constitutive relation \cite{madelung2004semiconductors}  
\begin{equation}
\mathbf{D} = \boldsymbol{\varepsilon}\mathbf{E}\label{eq:13}
\end{equation}
governs the polarization-dependent response, where $\mathbf{D} = [D_x,D_y,D_z]$ is the electric displacement vector and $\mathbf{E} = [E_x,E_y,E_z]$ is the electric field. For $p$ polarized waves, $\mathbf{E} = [E_x,0,E_z]$, and substitution yields
\begin{equation}
D_x = \varepsilon_{xx}E_x + \varepsilon_{xz}E_z,\label{eq:14}
\end{equation}
\begin{equation}
D_z = \varepsilon_{zx}E_x + \varepsilon_{zz}E_z. \label{eq:15}
\end{equation}
The off-diagonal terms $\varepsilon_{xz}$ and $\varepsilon_{zx}$ couple $E_x$ and $E_z$, and since their sign changes with propagation reversal, this coupling produces nonreciprocal reflectivity for $p$-polarized waves. In contrast, for $s$-polarized waves the electric field lies solely along the $y$-direction, $\mathbf{E} = [0,E_y,0]$, leading to
\begin{equation}
D_y = \varepsilon_{yy}E_y. \label{eq:16}
\end{equation}
Here, only the diagonal term $\varepsilon_{yy}$ contributes, and no off-diagonal coupling occurs. As a result, $s$-polarized waves remain reciprocal in a purely Voigt configuration. However, modest $s$-polarized nonreciprocity can arise in a hybrid MOKE configuration ($0^\circ<\varphi<90^\circ$), where the magnetization has both longitudinal and transverse components. In this case, the displacement field involves both diagonal and off-diagonal permittivity terms, enabling nonreciprocal effects. Although dual-polarized nonreciprocity can thus be observed, the effect remains small in single layer structures \cite{fang2023dual}.  

 The magnetization in each layer can be aligned in an arbitrary direction corresponding to a generalized hybrid MOKE configuration. Although the applied magnetic field orientation in each layer can be tuned through rotating the plane of incidence, the overall nonreciprocal contrast does not follow a simple vector summation of the individual layer contributions, since it depends on the interference between multiple reflected and transmitted fields, especially in the presence of polarization conversion. For example,  An incident wave first interacts with the top layer ($0^\circ < \varphi < 90^\circ$), where it is partially converted into $s$- and $p$-polarized waves with distinct phase and amplitude. As the wave propagates, each subsequent layer applies a different hybrid MOKE interaction due to varying polar angles, azimuthal orientations, and magnetization directions. The integrated contribution of all layers thus enhances both $s$- and $p$-polarized nonreciprocity. In the following, we first study the nonreciprocity of a 1-layer InAs structure when varying $\varphi$. We then consider a simple 2-layer InAs structure to show the feasibility of achieving dual-polarized nonreciprocity through layer-specific magnetization control. Then, to achieve strong broadband dual-polarized nonreciprocity, we design a gradient-doped multilayer heterostructure and tune the magnetization in each layer.

\subsection{1-layer InAs structure}

\begin{figure*}[t]
	\centering
	\includegraphics[width=\linewidth]{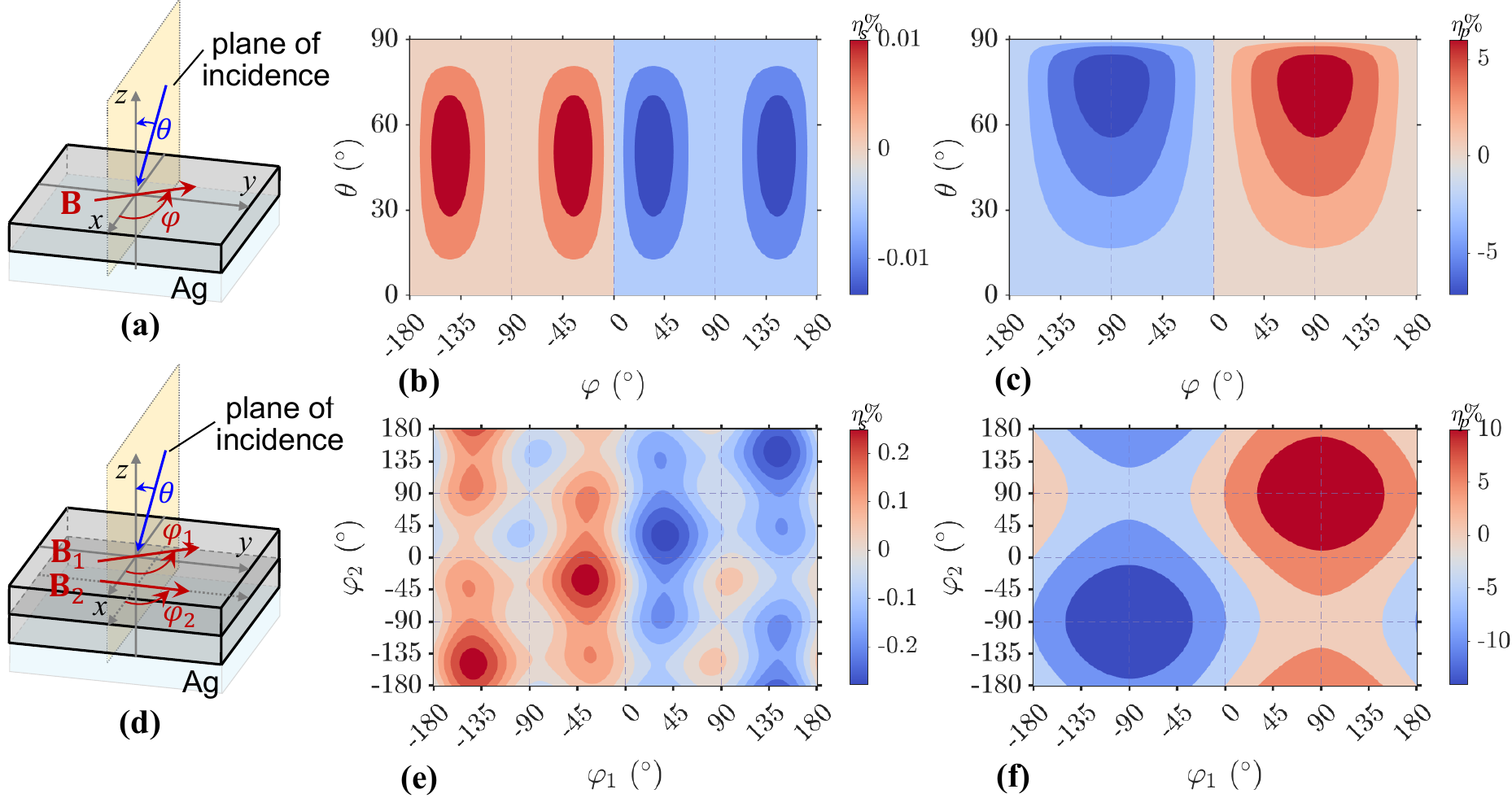}
\caption{
Polarization-independent contrasts for one- and  two-layer InAs structures. 
(a) Schematic of the one-layer InAs structure with doping level 
$n_{e} = 3.5 \times 10^{17}\,\mathrm{atoms/cm}^3$ and the normalized contrast for 
(b) $s$ and (c) $p$ polarizations. 
(d) Two-layer InAs structure and the corresponding normalized contrast for 
(e) $s$ and (f) $p$ polarizations. 
The doping levels for the first and the second layers are 
$n_{e,1} = 3.5 \times 10^{17}\,\mathrm{atoms/cm}^3$ and 
$n_{e,2} = 5.5 \times 10^{17}\,\mathrm{atoms/cm}^3$, respectively. Calculations are done at $\lambda=30 $ $\mu m$ and $\phi=0^{\circ}$. Polar angle is fixed as
$\theta=45^{\circ}$ for two-layer structure.}
\label{fig:Symmetry1}
\end{figure*}

{For the 1-layer InAs structure (\textbf{Figure}~\ref{fig:Symmetry1}(a)), we use the following fixed material parameters for InAs ~\cite{Do2025}: 
the high-frequency permittivity $\varepsilon_\infty = 12.37$, 
the effective electron mass $m^* = 0.033 m_e$, 
the damping rate $\Gamma = 5.9 \times 10^{12} \, \mathrm{rad/s}$, 
and the carrier concentration $n_e = 3.5 \times 10^{17} \, \mathrm{cm^{-3}}$. 
The layer thickness is $t = 1200 \, \mathrm{nm}$. 
As shown in \Cref{fig:Symmetry1}(b), the nonreciprocity for $s$-polarized waves is maximized when $\varphi = 45^{\circ}$. 
This behavior can be understood in terms of the conventional MOKE~\cite{qiu2000surface}. 
For $s$-polarized waves, nonreciprocity arises when %
$0^{\circ} < \varphi < 90^{\circ}$. 
Varying $\varphi$ effectively rotates the plane of incidence, modifying the off-diagonal components of the permittivity tensor and enabling polarization conversion. 
Consequently, an $s$-polarized wave can be partially converted into a $p$-polarized wave. 
Since $s$-wave nonreciprocity is strongly tied to cross-polarized reflection, its contrast is maximized when $\varepsilon_{xz}$ reaches its largest value, which occurs at $\varphi = 45^{\circ}$~\cite{Wu_2018}.  Although azimuthal tuning makes it possible to achieve dual-polarized nonreciprocity in a single magneto-optical layer, it comes at the cost of significantly reduced magnitude. The contrast vanishes at $\varphi = 0^{\circ}$, revealing an intrinsic angular restriction of single-layer designs (\Cref{fig:Symmetry1}). The polarization conversion responsible for $s$-wave nonreciprocity is exactly zero at $\varphi = 0^{\circ}$ and $\varphi = 90^{\circ}$, increasing monotonically to its maximum at $\varphi = 45^{\circ}$. For $p$-polarized waves (\Cref{fig:Symmetry1}(c)), nonreciprocity is largest when the magnetic field is in the Voigt configuration and the polar angle is highly oblique; in this case, the contrast increases from $\varphi = 0^{\circ}$ to $\varphi = 90^{\circ}$ and then decreases to zero at $\varphi = 180^{\circ}$.

These observations highlight the fundamental limitation of single-layer emitters. They inherently produce a narrowband and strongly angle-restricted nonreciprocal response that is maximized only for one polarization at a time. In contrast, multilayer designs with gradient doping and independently tuned magnetization orientations can break these constraints due to coupling between polarization conversions. Therefore, we next examine a two-layer structure to illustrate the key physical mechanisms, and then generalize these insights to the optimized multilayer geometry.

\subsection{2-layer InAs structure}
For the 2-layer InAs structure (\Cref{fig:Symmetry1}(d)), we treat the following material parameters as fixed \cite{Do2025}: $\varepsilon_\infty = 12.37$, $m^* = 0.033 m_e$, $\Gamma = 5.9 \times 10^{12} \, \mathrm{rad/s}$, and the carrier concentrations of the first and second layers $n_{e,1}=3.5 \times 10^{17} \, \mathrm{atoms/cm}^3$ and $n_{e,2}=5.5 \times 10^{17} \, \mathrm{atoms/cm}^3$, respectively.
The thicknesses of the two InAs layers are $t_1 = t_2 = 1200 \,\mathrm{nm}$.
The azimuthal and polar angles are fixed at $\phi=0^{\circ}$ and $\theta=45^{\circ}$.

In this structure, the $s$-polarized wave entering the first layer (hybrid MOKE configuration) is partially converted to a $p$-polarized wave. The $p$-polarized component interacts strongly with ENZ modes, which are inherently nonreciprocal. At the exit, the second layer enables partial conversion back to an $s$-polarized wave. $p$-polarized nonreciprocity is maximized when the magnetic fields in both layers are aligned in the Voigt configuration ($\varphi_{1,2} = 90^{\circ}$) where polarization conversion does not occur (\Cref{fig:Symmetry1}(f)). Compared with a single-layer emitter, the multilayer configuration enables substantially stronger and broadband nonreciprocity for both polarizations over a wider range of oblique polar angles thereby providing a more general and robust platform for high-performance thermal emitters. For example, in a single-layer structure (\Cref{fig:Symmetry1}(a-c))}, the $p$-polarized contrast barely reaches $2\%$, and the $s$-polarized contrast remains as low as $0.01\%$ ($\varphi=45^{\circ}$, $\theta=45^{\circ}$). In contrast, a two-layer structure in \Cref{fig:Symmetry1}(d-f) readily achieves approximately $0.2\%$ and $10\%$ for the $s$- and $p$-polarized cases, respectively ($\varphi_1,_2=45^{\circ}$, $\theta=45^{\circ}$).

These observations highlight that the overall nonreciprocal performance depends sensitively on the relative orientation of magnetization and the coupling between polarization states. Achieving a desired polarization-independent nonreciprocal emitter therefore requires simultaneously tuning a large number of parameters, including the magnetization, carrier concentration (Or Fermi level), and geometric layer properties. In the following section, we propose an the optimization algorithm used to tune these parameters which could be challenging to achieve.

\section{Optimization of polarization-independent contrasts}

Before presenting the optimization problem formulated for a structure of interest, we define the following normalized directional total contrasts: 
\begin{equation}\label{eqn:localcontrast}
	\eta_j({\bf x},\theta,\phi) = \frac{\int_{\lambda_\mathrm{L}}^{\lambda_\mathrm{U}} \eta_j({\bf x},\theta,\phi,\lambda) \diff \lambda}{\int_{\lambda_\mathrm{L}}^{\lambda_\mathrm{U}} 1 \diff \lambda}, \ \ j\in \{p,s\},
\end{equation}
where $\eta_j$ is the $j$-polarized spectral directional contrast given in \Cref{eq:7}. ${\bf x} = \left[\boldsymbol{\varphi}, {\bf p} \right]$ with $\boldsymbol{\varphi}$ representing the design variables to be optimized, which are the layer magnetization directions in this study, and ${\bf p}$ representing other known parameters of the structure.
$\lambda_\mathrm{L}$ and $\lambda_\mathrm{U}$ are the lower and upper values of the wavelength range considered, respectively.
1 in the denominator is the maximum possible value of $\eta_{j}({\bf x},\theta,\phi,\lambda)$. 
Note that, although the magnetization strength, carrier concentration, and geometric layer properties are treated as known parameters in this work, they could also be considered design variables without any change in the formulation of the optimization problem and the solution strategy.
Since ${\bf p}$, $\theta$, and $\phi$ are known parameters, we use $\eta_j(\boldsymbol{\varphi})$ hereafter to represent $\eta_j({\bf x},\theta,\phi)$.

\begin{figure*}[t]
	\centering
	\includegraphics[scale=0.55]{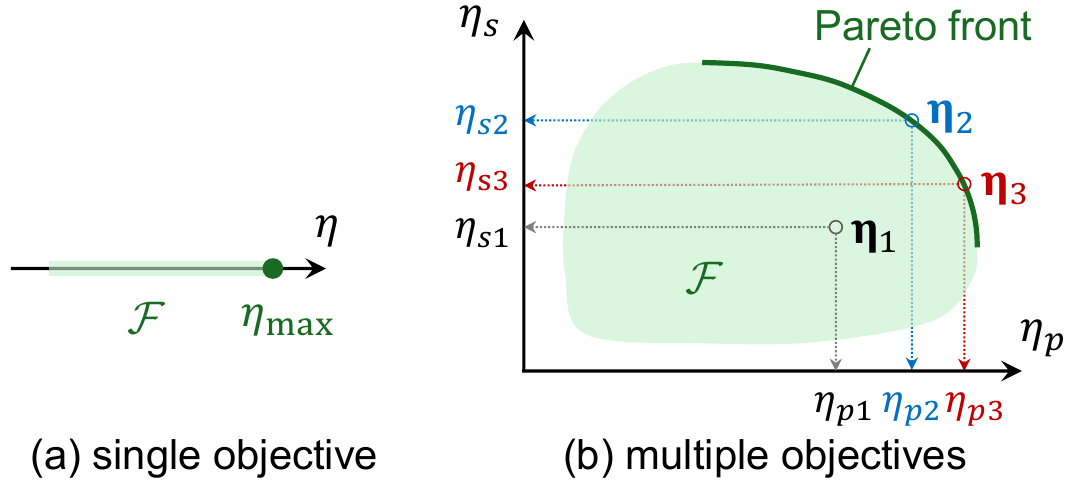}
	\caption{The objective spaces and solutions for single- and bi-objective maximization problems.}
	\label{fig:Pareto}
\end{figure*}

To achieve high-performance, dual-polarized nonreciprocal emitters using the pattern-free heterostructures in \Cref{fig:EmittersStruct}(b), we formulate the following bi-objective maximization problem:
\begin{equation}\label{eqn:moproblem}
	\underset{\boldsymbol{\varphi} \in \Phi}{\max} \ \  \left[ \eta_p(\boldsymbol{\varphi}), \eta_s(\boldsymbol{\varphi})\right],
\end{equation}
where $\Phi$ is the set of all possible values of $\boldsymbol{\varphi}$.
Unlike the single-objective maximization problem as illustrated in \textbf{Figure}~\ref{fig:Pareto}(a), which often has a unique optimal solution $\eta_{\max}$, problem (\ref{eqn:moproblem}) generally does not admit a unique optimal solution.
This is because both objective functions $\eta_p$ and $\eta_s$ do not reach their maximum at the same value of $\boldsymbol{\varphi}$. Instead, solutions to problem (\ref{eqn:moproblem}) are sought where no objective can be improved without worsening another.

We now formally define the solutions to problem (\ref{eqn:moproblem}).
Let $\boldsymbol{\varphi}_1, \boldsymbol{\varphi}_2 \in \Phi$ represent two feasible candidate solutions, and $\boldsymbol{\eta}_1  =[\eta_{p1},\eta_{s1}]=[\eta_p(\boldsymbol{\varphi}_1), \eta_s(\boldsymbol{\varphi}_1)]$ and $\boldsymbol{\eta}_2  =[\eta_{p2},\eta_{s2}]=[\eta_p(\boldsymbol{\varphi}_2), \eta_s(\boldsymbol{\varphi}_2)]$ the corresponding points in the objective space. $\boldsymbol{\varphi}_2$ is said to dominate $\boldsymbol{\varphi}_1$ (i.e., $\boldsymbol{\eta}_2$ dominates $\boldsymbol{\eta}_1$) if and only if $\eta_{j2} \geq \eta_{j1}$, $\forall j \in \{p,s\}$ and $\eta_{j2} > \eta_{j1}$, $\exists  j \in \{p,s\}$ \cite{Kochenderfer2019}.
A solution $\boldsymbol{\varphi} \in \Phi$ to problem (\ref{eqn:moproblem}) is called a Pareto optimal solution if there does not exist $\boldsymbol{\varphi}' \in \Phi$ such that $\boldsymbol{\varphi}'$ dominates $\boldsymbol{\varphi}$.  
As a result, there may exist many Pareto optimal solutions to problem (\ref{eqn:moproblem}). These solutions form the so-called Pareto set.
The image of the Pareto set in the objective space is called the Pareto front.

For example, consider three distinct points $\boldsymbol{\eta}_1$, $\boldsymbol{\eta}_2$, and $\boldsymbol{\eta}_3$ in the space $\mathcal{F}$ of $\eta_p$ and $\eta_s$, as illustrated in \Cref{fig:Pareto}(b). 
Since $\eta_{j2} > \eta_{j1}$ and $\eta_{j3} > \eta_{j1}$, $\forall j \in \{p,s\}$, both $\boldsymbol{\eta}_2$ and $\boldsymbol{\eta}_3$ dominate $\boldsymbol{\eta}_1$. 
Neither $\boldsymbol{\eta}_2$ nor $\boldsymbol{\eta}_3$ dominates the other because neither point is strictly better than the other in all objectives.
Since $\boldsymbol{\eta}_2$ and $\boldsymbol{\eta}_3$ are not dominated by any other point in $\mathcal{F}$, they are two members of the Pareto front.  

\section{Optimization solver}
There are two primary strategies to find Pareto optimal solutions for a multi-objective optimization problem: decomposition and population methods.
The decomposition methods convert the multi-objective problem to a series of single-objective problems, typically using constraint- or weight-based approaches \cite{Haimes1971,Das1997}.
The former constrains all but one of the objectives and generates members of the Pareto front by varying the constraint values.
The latter encodes preferences among the objectives using a weight vector and aggregates them into a single objective defined as the weighted sum of the objectives. It then traces the Pareto front by sweeping the weights over the space of their possible values.
The population methods, starting from an initial population of candidate solutions, incrementally improve the diversity of individuals that span the Pareto front. 
One of the notable population multi-objective algorithms is the non-dominated sorting genetic algorithm II (NSGA-II) \cite{Deb2002} that is used in this study.

NSGA-II encodes each design point $\boldsymbol{\varphi}$, referred to as an individual of a set of many individuals called a population, as a chromosome that is typically a binary string in its simplest form \cite{Kochenderfer2019}. At each generation (i.e., optimization iteration), individuals with better objective function values are expected to be selected to propagate their genetic material (i.e., chromosomes of bits) to the next generation through the application of two genetic operators: crossover and mutation. The former combines the chromosomes of individuals to form new individuals called children, while the latter flips several bits in the chromosomes of a small number of randomly selected individuals.
The main steps of NSGA-II are summarized as follows:
\begin{itemize}
    \item {Step 1:} Randomly generate a population of size $N$ of design variables $\boldsymbol{\varphi}$.
    
    \item {Step 2:} Evaluate the objective functions for all individuals in the population.

    \item {Step 3:} Rank the individuals into different Pareto front levels based on their objective values and the dominance concept. Specifically, level 1 consists of non-dominated individuals in the population, level 2 includes non-dominated individuals except those in level 1, and so on.

    \item {Step 4:} Compute the so-called crowding distance for each individual, which is the average distance of two points on either side of this point along each of the objectives \cite{Deb2002}. This step is to ensure a well-distributed set of solutions across the objective space.

    \item {Step 5:} Select parents for the genetic operators based on the nondomination ranks and crowding distances. Specifically, when comparing two candidate solutions with different nondomination ranks, the one with the lower (i.e., better) rank is preferred. If both solutions belong to the same front level, preference is given to the solution residing in a less crowded region of the objective space to promote the population's diversity \cite{Deb2002}.

    \item {Step 6:} Apply crossover and mutation operators to generate children. Evaluate the objective functions for all generated children.

    \item {Step 7:} Combine the parent and children populations, sort the combined population, and select the top $N$ individuals for the next generation.

    \item {Step 8:} Repeat Steps 3 to 7 until a termination criterion is met (e.g., number of optimization generations).
\end{itemize}

When the objective functions are expensive to evaluate,
multiobjective Bayesian optimization methods can be used
to reduce the evaluation cost significantly \cite{Do2025edu, Do2025mfbo}.
We do not use these methods in this study because the directional contrasts,
given all the known parameters, can be computed inexpensively.

\section{Symmetry of the nonreciprocal contrast}
\label{symmetry}
Since we consider the sign of $\eta_j(\boldsymbol{\varphi})$, we can obtain optimal configurations with negative $\eta_j(\boldsymbol{\varphi})$ values.
This raises a question of how we can still use these configurations to propose new configurations that have strong positive contrasts.
To address this, we examine the symmetry and antisymmetry of $\eta_j(\boldsymbol{\varphi})$.

{We consider 1- and 2-layer InAs structures, as shown in \textbf{Figures}~\ref{fig:Symmetry1}(a) and (d), respectively.
For 1-layer InAs structure, we set $\phi=0^{\circ}$ (i.e., the plane of incidence is the $xz$ plane) while varying $\theta$ and $\varphi$.
For 2-layer InAs structure, we set $\phi=0^{\circ}$ and $\theta=45^{\circ}$ while varying $\boldsymbol{\varphi}=[\varphi_1,\varphi_2]$, where $\varphi_1$ and $\varphi_2$ are the magnetization directions in the first and second layers, respectively.
The resulting contrasts $\eta_j(\boldsymbol{\varphi})$, plotted in Figures~\ref{fig:Symmetry1}(b), (d), (e), and (f), reveal the symmetry and antisymmetry in $\eta_j(\boldsymbol{\varphi})$ across the two considered structures. Below, we provide an explanation for these observations.

As outlined in previous studies \cite{GuoC2022}, thermal emitters can be classified into three categories based on their Shubnikov point groups \cite{Dresselhaus_2008, bradley2009mathematical}: gray, colorless, and black-and-white groups. Gray group emitters consist of fully reciprocal media where Kirchhoff’s law strictly applies, ensuring that directional spectral emissivity and absorptivity are always equal. In contrast, colorless group emitters are nonreciprocal media in which no direct relationship between emissivity and absorptivity exists, aside from constraints imposed by global energy conservation. Black-and-white group emitters are also nonreciprocal systems but possess geometrical symmetry-imposed constraints that enforce modified forms of Kirchhoff’s law, linking specific components of emissivity and absorptivity components. Since we employ nonreciprocal materials such as InAs and magnetic Weyl semimetals with planar structure who holds geometrical symmetry, they belong to black-and-white groups. Importantly, while these materials exhibit nonreciprocity, they do not necessarily break all symmetry. In particular, the magnetic Weyl semimetals used in our study are not spatially chiral, meaning that certain symmetry operations are still preserved. 

This symmetry constraint leads to the adjoint Kirchhoff’s law for black and white group emitters, which is enforced \Cref{eq:3} as introduced earlier \cite{GuoC2022}. Consequently \Cref{eq:1} directly yields 
\begin{equation}
\eta_j (\theta,\phi,\lambda) = - \eta_j (\theta,-\phi,\lambda)
\label{eq:eta_antisym}.
\end{equation}
indicating that the nonreciprocity is anti-symmetric with respect to the azimuthal angle $\phi$ (for fixed $\varphi$, $\theta$, and $\lambda$). This anti-symmetry originates from the underlying spatial symmetry of planar nonreciprocal systems. Importantly, this behavior can be reproduced equivalently by fixing the azimuthal angle $\phi$ and instead rotating the applied magnetic field $\varphi$. Because the nonreciprocity depends only on the relative angle between the plane of incidence and the applied magnetic field, variations in $\phi$ and corresponding variations in $\varphi$ are interchangeable. Thus,

\begin{equation}
\eta_j (\boldsymbol{\varphi}) = -\eta_j (-\boldsymbol{\varphi})
\label{eq:eta_sym}.
\end{equation}
This justifies the behavior observed in Figures~\ref{fig:Symmetry1}(b) and (c). For multilayer heterostructures, the same anti-symmetry constraint applies to group-wise rotations of the magnetization directions $\boldsymbol{\varphi}$, as demonstrated in Figures~\ref{fig:Symmetry1}(e) and (f). 

\section{Optimization results and discussion}

\subsection{2-layer InAs structure}

{We now find the Pareto solutions for the 2-layer InAs structure in \Cref{fig:Symmetry1}(d) by solving problem~(\ref{eqn:moproblem}) with $\varphi_1$ and $\varphi_2$ varying within $[-180^{\circ}, 180^{\circ}]$. We perform two independent NSGA-II trials to examine how their inherent randomness influences the final Pareto solutions. Each NSGA-II trial uses a population size of $1000$, a crossover fraction of $0.65$, and a function tolerance of $10^{-12}$, and is run for $100$ generations. As shown in \textbf{Figure}~\ref{fig:optsolution_2InAs}(a), both NSGA-II trials yield identical Pareto fronts, where the maximum $\eta_s$ is much smaller than the maximum $\eta_p$. The Pareto optimal solutions represent how we can prioritize the design objectives. For example, if the goal is to maximize $\eta_p$, achieve zero value of $\eta_p$, or maximize $\eta_s$, then solution (1), (2), or (3), as marked in \Cref{fig:optsolution_2InAs}(a), can be selected, respectively. Interestingly, solution (1) is also optimal when the goal is to maximize the unpolarized contrast $(\eta_p+\eta_s)/2$.

\begin{figure*}[t]
	\centering
	\includegraphics[width=\linewidth]{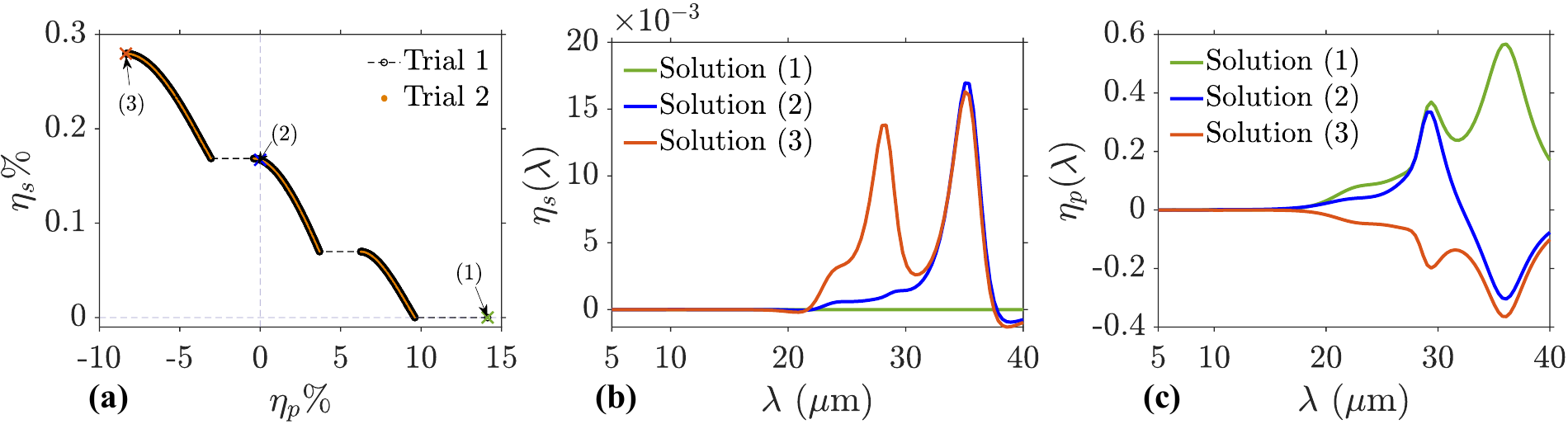}
	\caption{Optimization results for the 2-layer InAs structure. (a) Pareto fronts from two optimization trials for the 2-layer InAs structure and three representative solutions: solution (1) where $\eta_p$ is maximized, solution (2) where $\eta_p$ is zero, and solution (3) where $\eta_s$ is maximized. (b), (c) Directional $s$- and $p$-polarized contrasts for solutions (1), (2), and (3), respectively. Polar angle is fixed as
$\theta=45^{\circ}$.}
	\label{fig:optsolution_2InAs}
\end{figure*}

\begin{table*}[t]
	\centering
	\caption{Solutions (1), (2), and (3) selected from the Pareto-optimal solutions of the 2-layer InAs structure (angles in degree unit).}
	\begin{tabular}{cccl}
            \hline\noalign{\smallskip}
		Solution & Bi-objective optimization & Single-objective optimization & Note\\
		\hline\noalign{\smallskip}
		$(1)$ & $[90.0, 90.0]$ & $[90.0, 90.0]$ & maximum $\eta_p$ \\
		$(2)$  & $[-32.3, 80.7]$ & -- & zero $\eta_p$\\
		$(3)$  & $[-37.0, -30.9]$, $[-143.0	-149.1]$ & $[-37.0	-30.9]$, $[-143.0	-149.1]$ & maximum $\eta_s$ \\
		\hline\noalign{\smallskip}
	\end{tabular}
	\label{table2InAs}
\end{table*}

\Cref{table2InAs} lists the optimal values of $\boldsymbol{\varphi} = [\varphi_1,\varphi_2]$ corresponding to solutions (1), (2), and (3). The corresponding directional nonreciprocity contrasts are presented in 
Figures.~\ref{fig:optsolution_2InAs}(a–c), enable a direct comparison of the 
three optimized configurations.

For solution (1), the optimal $\boldsymbol{\varphi}$ aligns the system 
in the Voigt configuration, where the external magnetic field is perpendicular to 
the plane of incidence. In this geometry, polarization is preserved, and consequently the 
$s$-polarized directional contrast vanishes identically while $\max\eta_p=10\%$. This behavior is fully consistent with MOKE constraints, as Voigt geometry supports nonreciprocal effects only for $p$-polarized waves.

Solution (2) exhibits more intricate behavior. Although the spectrally 
integrated $p$-polarized contrast $\eta_p$ evaluates to zero over the wavelength range $5~\mu\mathrm{m}$ to $40~\mu\mathrm{m}$ (Figure \ref{fig:optsolution_2InAs}(a)), the spectrally 
resolved contrast reveals two equal but oppositely signed peaks around $30~\mu\mathrm{m}$ and $35~\mu\mathrm{m}$ ($\eta_p=\pm0.3$). These contributions cancel in the integrated value, yet the opposite signs provides tunable control over the sign of nonreciprocity in different spectral regions. Moreover, these peaks correspond to the ENZ responses of the second and first layers, associated with $n_{e,2}$ and $n_{e,1}$, respectively, indicating that gradient doping in multilayer structures can support broadband and spectrally reconfigurable nonreciprocity. The results also highlight the importance of considering a broad spectral range in designing structures for energy and thermal applications.

Since the optimal configuration for solution (2) satisfies 
$\boldsymbol{\varphi}\neq0^{\circ},90^{\circ}$, polarization conversion occurs, enabling a 
nonzero $s$-polarized contrast $\max\eta_s=0.017$ (Figure~\ref{fig:optsolution_2InAs}(b)). Thus, this configuration exhibits 
dual-polarized nonreciprocity within specific spectral windows, even though the 
integrated $p$-polarized nonreciprocity integrates to zero across the full 
wavelength domain.

For solution (3), two distinct sets of optimal $\boldsymbol{\varphi}$ 
are identified, each yielding the maximum $s$-polarized contrast. The existence 
of these two equivalent solutions follows directly from the symmetry relation 
discussed in \Cref{symmetry}. As shown in 
Figure~\ref{fig:optsolution_2InAs}(a), the maximum $s$-polarized contrast can 
reach values as $0.28\%$, while the $p$-polarized contrast remains as high as $8\%$. Moreover, the fact that the $s$- and $p$-polarized contrasts possess opposite 
signs provides an additional degree of freedom to control the direction of nonreciprocity independently for each polarization channel.

To further validate solutions (1) and (3), which are selected from the Pareto set, we formulate and solve two single-objective maximization problems: one targeting the maximization of $\eta_p$ and the other targeting the maximization of $\eta_s$.
The resulting solutions reported in the third column of \Cref{table2InAs} validate solutions (1) and (3). Overall, these results demonstrate that the proposed heterostructure can be 
optimized not only to maximize directional contrast for a target polarization, 
but also to realize polarization-independent or dual-polarization nonreciprocal 
thermal emission within the same design framework.

\subsection{6-layer 3InAs+3Weyl structure}
Although the 2-layer InAs structure shows the feasibility of dual-polarized nonreciprocal effect by controlling layer-specific magnetizations, this effect is still modest. To achieve a stronger dual-polarized nonreciprocity, we design a six-layer structure composed of three gradient-doped InAs layers atop three gradient ENZ layers of magnetic Weyl semimetals (\textbf{Figure}~\ref{fig:optsolution_3InAs3Weyl}(a)). In our earlier work \cite{Do2025}, a similar six-layer platform was optimized exclusively for $p$-polarized nonreciprocity, where polarization conversion was intentionally suppressed by fixing the magnetization direction in a Voigt configuration. That approach enabled an optimized nonreciprocal emitter for a single polarization channel. In contrast, in the present study we allow the magnetization direction in each layer to rotate freely, fully incorporating polarization-conversion effects into the design space.  Because this expanded design requires simultaneous tuning of multiple magnetization vectors and material parameters, we employ a multi-objective optimization strategy based on Pareto fronts. %
This broader framework enables the identification of structures that maximize dual-polarized nonreciprocity.

The thickness values for the first three InAs layers are $t_1 = t_2 = t_3 = 1200 \, \mathrm{nm}$, and those of the last three Weyl layers are $t_4 = t_5 = t_6 = 200 \, \mathrm{nm}$.
The carrier concentrations of the first, second, and third InAs layers are $n_{e,1}=3.5 \times 10^{17}$, $n_{e,2}=4.5 \times 10^{17}$,  $n_{e,3}=5.5 \times 10^{17} \, \mathrm{atoms/cm}^3$.
The high-frequency permittivity, effective electron mass, and damping rate for the InAs layers are the same as those of the 2-layer InAs configuration.
For the three Weyl layers, the background permittivity, number of Weyl points, Fermi velocity, relaxation time, and cut-off energy of Weyl layers are $\varepsilon_b = 6.2$, $g=2$, $\tau = 10^{-12}\,\mathrm{s}$, $\nu_F = 0.83 \times 10^5 \, \mathrm{m/s}$, and $E_C = 0.45 \, \mathrm{eV}$, respectively \cite{ZhaoB2020}.
The temperature is fixed at $T=300 \, \mathrm{K}$.
The Fermi levels of the three Weyl layers, from top to bottom, are set to $E_{F,1} = E_{F,2} = E_{F,3} = 0.05 \, \mathrm{eV}$.
Thus, there are six design variables for this structure, including the layer magnetization directions $\varphi_1,\dots, \varphi_6$. These variables are selected from $[-180^{\circ},180^{\circ}]$.

We perform two independent NSGA-II trials, each with a population size of 2000. Other parameters are identical to those of the NSGA-II trials for the 2-layer InAs structure.

\begin{figure*}[t]
	\centering
	\includegraphics[scale=0.70]{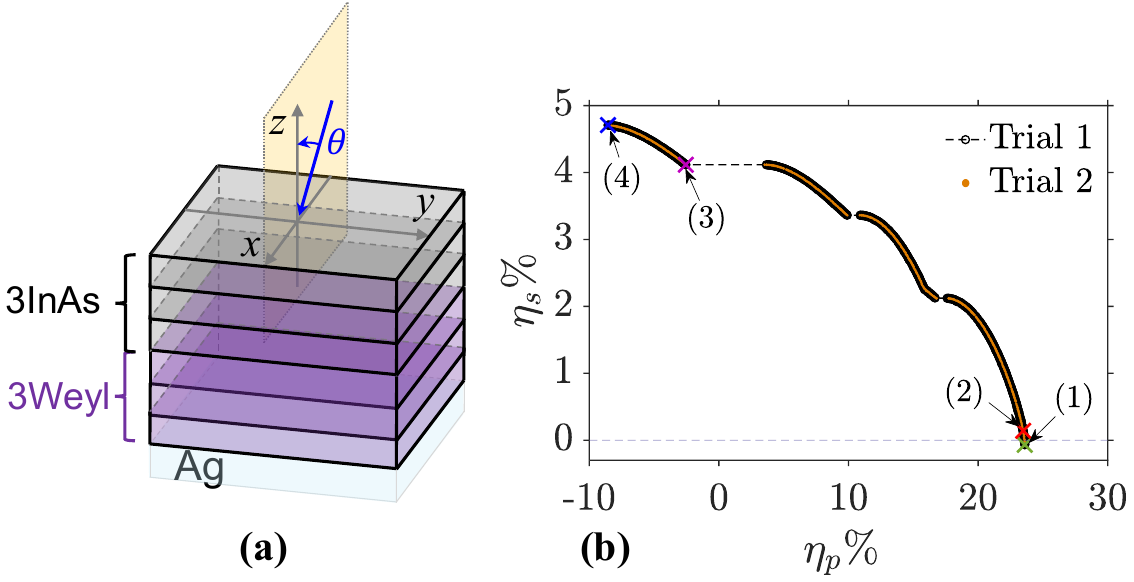}
	\caption{(a) Schematic of the 6-layer 3InAs+3Weyl structure. (b) Pareto fronts from two optimization trials. Solutions (1), (2), (3), and (4) from the Pareto front correspond to the configurations at which $\eta_p$ is maximized, $(\eta_p+\eta_s)/2$ is maximized, $\eta_p$ is closest to zero, and $\eta_s$ is maximized, respectively. Polar angle is fixed as
$\theta=45^{\circ}$.}
	\label{fig:optsolution_3InAs3Weyl}
\end{figure*}

\begin{table*}[t]
	\centering
	\caption{Solutions (1), (2), (3), and (4) selected from the Pareto-optimal solutions of the 6-layer 3InAs+3Weyl structure (angles in degree unit).}
    \setlength{\tabcolsep}{2pt} %
	\begin{tabular}{cccl}
        \hline\noalign{\smallskip}
		Solution & Bi-objective optimization & Single-objective optimization & Note\\
		\hline\noalign{\smallskip}
		$(1)$ & {\small$[91.7, 87.0, 86.8, 158.7, 19.5, 38.4]$} & {\small$[89.3, 88.6, 85.4, 159.0, 19.3, 40.9]$} & maximum $\eta_p$\\
		$(2)$  & {\small$[92.1,	86.8, 77.9, 153.7, 12.7, 30.1]$} & {\small$[90.2, 87.6, 78.2, 153.4, 12.0, 30.5]$} & maximum $(\eta_p+\eta_s)/2$\\
		$(3)$  & {\small$[-13.7, -15.6, 21.3,	-42.0, -30.2, 59.4]$} & -- & closest-zero $\eta_p$\\
        $(4)$  & {\small$[-37.7, -32.2, -10.9, -43.7, -31.2, 50.3]$} & {\small$[-38.9,	-33.3,	-13.0, -44.5, -31.8, 46.6]$} & maximum $\eta_s$\\
		\hline\noalign{\smallskip}
	\end{tabular}
	\label{table2}
\end{table*}

\begin{figure*}[t]
	\centering
	\includegraphics[scale=0.7]{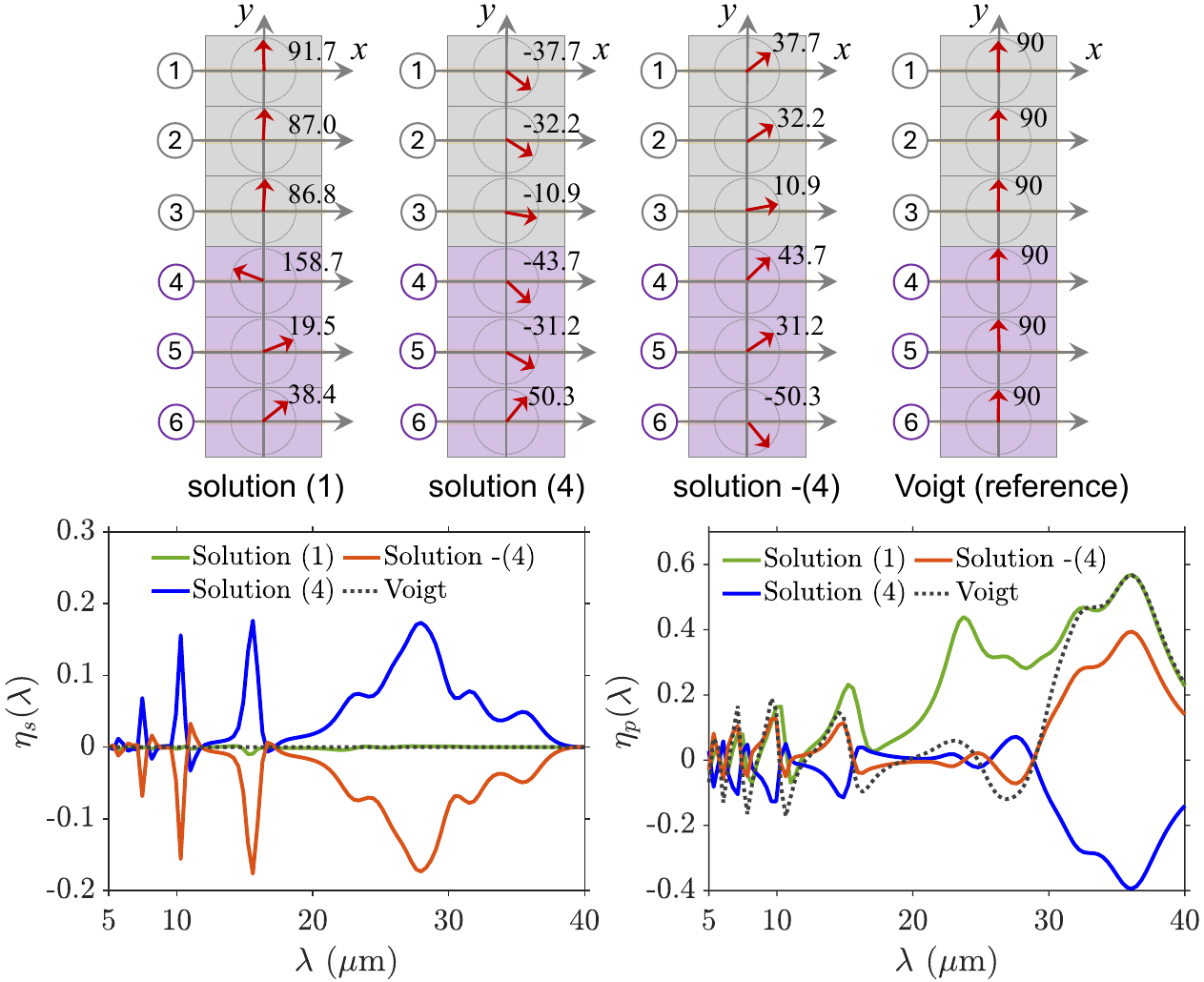}
	\caption{Configurations and directional contrasts for several solutions of the 6-layer 3InAs+3Weyl structure. Solution (1) with maximum $\eta_p$, solution (4) with maximum $\eta_s$, solution -(4) obtained by flipping $\boldsymbol{\varphi}$ of solution (4) about $x$ axis, and Voigt configuration as reference.}
	\label{fig:optcontrast_3InAs3Weyl}
\end{figure*}

\Cref{fig:optsolution_3InAs3Weyl}(b) shows four representative solutions selected from the Pareto front. Specifically, solutions (1), (2), (3), and (4) correspond to the configurations at which $\eta_p$ is maximized, $(\eta_p+\eta_s)/2$ is maximized, $\eta_p$ is closest to zero, and $\eta_s$ is maximized, respectively.
The corresponding optimal angles $\boldsymbol{\varphi}=\left[ \varphi_1,\dots, \varphi_6 \right]$ are listed in the second column of \cref{table2}, and are consistent with those obtained from solving the respective single-objective maximization problems.
\textbf{Figure}~\ref{fig:optcontrast_3InAs3Weyl} shows configurations of solutions (1), (4), 
and -(4) along with their corresponding directional contrasts. For comparison, 
the directional contrasts of the Voigt configuration reported in our previous 
work \cite{Do2025} are also included. The resulting Pareto fronts are nearly identical, as shown in \Cref{fig:optsolution_3InAs3Weyl}(b). None of the Pareto-optimal solutions exhibit zero $\eta_p$ when both $\eta_p$ and $\eta_s$ are simultaneously optimized. This, however, does not imply that solutions with zero $\eta_p$ cannot be found.

For solution (1) it is observed that $\max\eta_p = 23.6\%$ is achieved 
when the external magnetic field applied to the InAs layers is oriented nearly 
perpendicular to the plane of incidence ((\textbf{Figure}~\ref{fig:optsolution_3InAs3Weyl}(b)). An interesting observation is that the maximum $\eta_p$ value from our previous study $(4.7\%)$ \cite{Do2025} is approximately $20\%$ of the maximum $\eta_p$ value $(23.6\%)$ in this work, indicating a notable improvement in nonreciprocal performance. This orientation largely suppresses 
polarization conversion in the first three layers. However, due to the 
magnetization directions in the Weyl layers and the resulting polarization 
coupling, the $p$-polarized nonreciprocity becomes significantly enhanced over a 
broader bandwidth compared to the Voigt configuration. Moreover, the 
$s$-polarized contrast shows improvement ($\max\eta_s=0.5\%$) in comparison to Voigt configuration which $s$-polrized contrast remains zero. 

In the shorter-wavelength region ($5-10~\mu\mathrm{m}$), the $p$-polarized 
contrast closely matches the values obtained in the Voigt configuration. 
At longer wavelengths ($10-40~\mu\mathrm{m}$), however, the contrast 
exceeds the Voigt configuration response, demonstrating a clear enhancement enabled by 
the optimized magnetization orientations. These results show that by employing a Pareto-front optimization framework, we 
can achieve nonreciprocity levels that surpass the previously established upper 
limit of the Voigt configuration, and importantly, do so across a substantially 
broader spectral bandwidth (\Cref{fig:optcontrast_3InAs3Weyl}). It should be mentioned that each peak in the 
contrast spectra corresponds to the ENZ frequencies of the InAs and Weyl layers which is responsible for broadband behavior similar to the 2-layer structure.

For solution (2), the optimization aims to enhance the unpolarized 
nonreciprocity. In this scenario, the response is primarily governed by the 
$p$-polarized channel, which exhibits a significant contrast of 
$\eta_p = 22\%$, while the $s$-polarized contribution is $\eta_s = 0.2\%$ (\cref{fig:optsolution_3InAs3Weyl}). This behavior is consistent with the results summarized in \cref{table2}. Specifically, $\boldsymbol{\varphi}$ obtained for solution (2)—which maximizes unpolarized nonreciprocity—closely matches the configuration identified in solution (1), where $\eta_p$ alone is maximized. This similarity confirms that the unpolarized performance in solution (2) is primarily driven by the $p$-polarized response.

When the objective is to suppress the $p$-polarized nonreciprocity, solution (3) 
achieves this requirement. Conventionally, such suppression can only be realized 
by aligning the sample in the longitudinal MOKE configuration. However, through 
the use of multilayer structure and appropriately 
oriented magnetization, we observe a near-zero $p$-polarized contrast 
$\eta_p$ in non-longtitudinal MOKE configuration, consistent with the results reported in \cref{table2}. At the same 
time, this configuration yields a non-negligible $s$-polarized nonreciprocity $\eta_s=3\%$, 
arising from the magnetic alignment and the MOKE.

In the configurations corresponding to solutions (4) and (-4), the magnetization angle satisfies $\varphi \neq 90^\circ$ in each layer (solution -(4) is obtained, according to the symmetry by flipping $\boldsymbol{\varphi}$ of solution (4) about $x$-axis.). Consequently, polarization conversion occurs, and the converted components interact with one another during transmission and reflection. This polarization coupling leads to a reduced $\eta_p$ compared with the Voigt configuration; however, it simultaneously yields a significantly improved $\eta_s$ (with $\max\eta_p = 0.38$ and $\max\eta_s = 0.17$).  Furthermore, difference between sign of $s$- and $p$-polarized contrast provides a shifting knob to control the sign of nonreciprocity. 

The optimal design of the thermal emitter strongly depends on the chosen nonreciprocity objective. While the solution (1) remains the most effective approach for maximizing $p$-polarized nonreciprocity, solutions (4) and (-4) provide the highest achievable $s$-polarized nonreciprocity. Thus, a nonreciprocal thermal emitter can be designed based on the targeted objective functions.

\section{Conclusion}
This work introduces texture-free multilayer heterostructures composed of InAs and magnetic Weyl semimetals as fabrication-friendly designs for achieving polarization-independent nonreciprocal thermal radiation. 
By leveraging the non-additive magneto-optical interactions, our designs overcome the limitations of conventional planar structures, enabling reciprocity breaking for both $p$- and $s$-polarizations without relying on surface patterning or nanoscale texturing.
Furthermore, by employing a Pareto optimization approach, we systematically identify layer-specific magnetizations for different configurations that exhibit intriguing dual-polarized nonreciprocal properties. Our study sheds light on designing dual-polarized nonreciprocal thermal emitters, establishing that $p$- and $s$-polarized contributions can be tuned according to the chosen objective functions, particularly given the typically stronger role of the $p$-polarized wave. Our results highlight that the physical mechanisms governing nonreciprocal responses in multilayer systems with rotated magnetization points to a promising direction for future theoretical development of dual-polarized nonreciprocal thermal emitters.

\medskip
\textbf{Acknowledgments} \par %
The authors acknowledge the funding from the University of Houston through the SEED program and the National Science Foundation, USA under Grant No. CBET-2314210, and the support of the Research Computing Data Core at the University of Houston, USA for assistance with the calculations carried out in this work.

\medskip

\bibliographystyle{unsrtnat}
\bibliography{ContrastsOptimization}

\end{document}